\documentclass[preprint,showpacs,amsmath,amssymb,prd]{revtex4}
\usepackage{dcolumn}
\usepackage{epsfig}
\usepackage{bm}
\begin{document}


\title{\large\bf
Discovery Potential of New Boson $W_1^{\pm}$ in the Minimal Higgsless Model at LHC}
\author{Jian-Guo~Bian$^{1}$}
\email{bianjg@mail.ihep.ac.cn}
\author{Guo-Ming~Chen$^1$}
\author{Ming-Shui~Chen$^1$}
\author{Zu-Hao~Li$^1$}
\author{Song~Liang$^1$}
\author{Xiang-Wei~Meng$^1$}
\author{Yong-Hui~Qi$^2$}
\author{Zhi-Cheng~Tang$^1$}
\author{Jun-Quan~Tao$^1$}
\author{Jian~Wang$^1$}
\author{Jian~Wang$^3$}
\author{Xian-You~Wang$^4$}
\author{Jian-Xiong~Wang$^1$}
\author{Hong~Xiao$^1$}
\author{Min~Yang$^1$}
\author{Jing-Jing~Zang$^1$}
\author{Zheng~Wang$^1$}
\author{Bin~Zhang$^2$}
\author{Zhen~Zhang$^1$}
\author{Zhen-Xia~Zhang$^1$}

\affiliation{\it
$^1$ Institute of High Energy Physics, Beijing 100049, China\\
$^2$ Center for High Energy Physics, Tsinghua University, Beijing 100084, China\\
$^3$ Physics College, Graduate University of Chinese Academy of Sciences, Beijing 100049, China\\
$^4$ Theoretical Physics Institute, ChongQing University, ChongQing 400044, China
}

\date{\today}

\begin{abstract}
In this paper, we demonstrate the LHC discovery potential of new charged vector
boson $W_1^{\pm}$ predicted by the Minimal Higgsless model in the process
$pp\rightarrow W_1^{\pm}qq^{\prime}\rightarrow W^{\pm}Z^0qq^\prime\rightarrow \ell^{\pm}\ell^+\ell^-\nu
qq^{\prime}(\ell=e,\mu)$
by analyzing  the generator level events of the signal and backgrounds.
The generator for the signal $pp\rightarrow {W_1}^{\pm}qq^\prime\rightarrow W^{\pm}Z^0qq^\prime$
at tree level is developed with the Minimal Higgsless model
and then interfaced with PYTHIA for the   
parton showers and hadronization.
The backgrounds are produced with PYTHIA and ACERMC.
We give  integrated luminosities required to discover
5$\sigma$ signal as a function of $W_1^{\pm}$ mass.
\end{abstract} 


\maketitle

\section{\noindent Introduction}

The Higgs boson of the electroweak standard model has been sought in vain by all known experiments
so far. The absence of such a Higgs boson will violate perturbative unitarity of the high energy
longitudinal weak boson scattering $V_LV_L \to V_LV_L$ ($V=W^\pm,~Z^0$).  To postpone this
unitarity violation, the Higgsless models are proposed, in which the new spin-1 gauge bosons
(rather than spin-0 Higgs scalars) play the key role in both 5d and 4d realizations$^{[1]}$.


Ref.[2] presented the first LHC-study on the distinct signatures of the Minimal Higgsless Model
which is exactly gauge-invariant and predicts just a pair of new gauge bosons $(W_1^\pm,~ Z_1^0)$
as light as about  400 GeV.
In ref.[2], the signature of new boson $W^{\pm}_1$  in the process
$pp\rightarrow W^{\pm}Z^0qq^\prime\rightarrow \nu3\ell qq^\prime$ and the process
$pp\rightarrow W^{\pm}Z^0Z^0\rightarrow qq^\prime 4\ell~(\ell=e,\mu)$
were investigated at the parton level, without
the initial and final state  parton showers and
hadronization for the signal and backgrounds,
while only the background  $pp\rightarrow
W^{\pm}Z^0qq^{\prime}$ was considered, which included non fusion and fusion processes.
Assuming $W^{\pm}_1$ mass within the region of
(550, 750)GeV,
the integrated luminosity required for $5\sigma$
$W^{\pm}_1$ signal detection is from about 
12/fb to 25/fb
for the first
channel, while from about 
40/fb to 300/fb 
for the second channel$^{[2]}$.

This work presents the results on the study of $W_1^\pm$ production
in the  process $pp\rightarrow W_1^{\pm}qq^\prime
\rightarrow W^{\pm}Z^0qq^\prime$
by analyzing the fully hadronized events of the signal and backgrounds
at the generator level.
The final state is chosen to be $\ell^\pm\nu \ell^+\ell^- qq^\prime~(\ell=e,\mu)$,
because electrons and muons are well identified by the LHC detectors.
The generator for the signal $pp\rightarrow W_1^\pm qq^\prime\rightarrow
W^\pm Z^0qq^\prime \rightarrow \ell^\pm\nu \ell^+\ell^- qq^\prime$
is developed based on the Minimal Higgsless model and corporated with PYTHIA$^{[3]}$ for the
parton showers and
parton hadronization. 

 Fig. 1 is the Feynman diagram for the process
 $pp\rightarrow W_1^{\pm}qq^{\prime}\rightarrow W^{\pm}Z^0qq^{\prime}\rightarrow \ell^{\pm}\ell^+\ell^-\nu qq^{\prime}$.
 The intermediate state $W_1^{\pm}$ proceeds through $W^{\pm}Z^0$ fusion. The typical cross sections (assuming
 $m_{W_1}$ to be 700 GeV) are 2.828 fb and 1.538 fb for 
 $pp\rightarrow W_1^+qq^{\prime}\rightarrow W^+Z^0qq^{\prime}\rightarrow \ell^+\nu\ell^+\ell^- qq^{\prime}$ and
 $pp\rightarrow W_1^-qq^{\prime}\rightarrow W^-Z^0qq^{\prime}\rightarrow \ell^-\nu\ell^+\ell^- qq^{\prime}$ respectively.


\begin{figure}[hbtp]
  \begin{center}
      \resizebox{6cm}{2.5cm}{\includegraphics{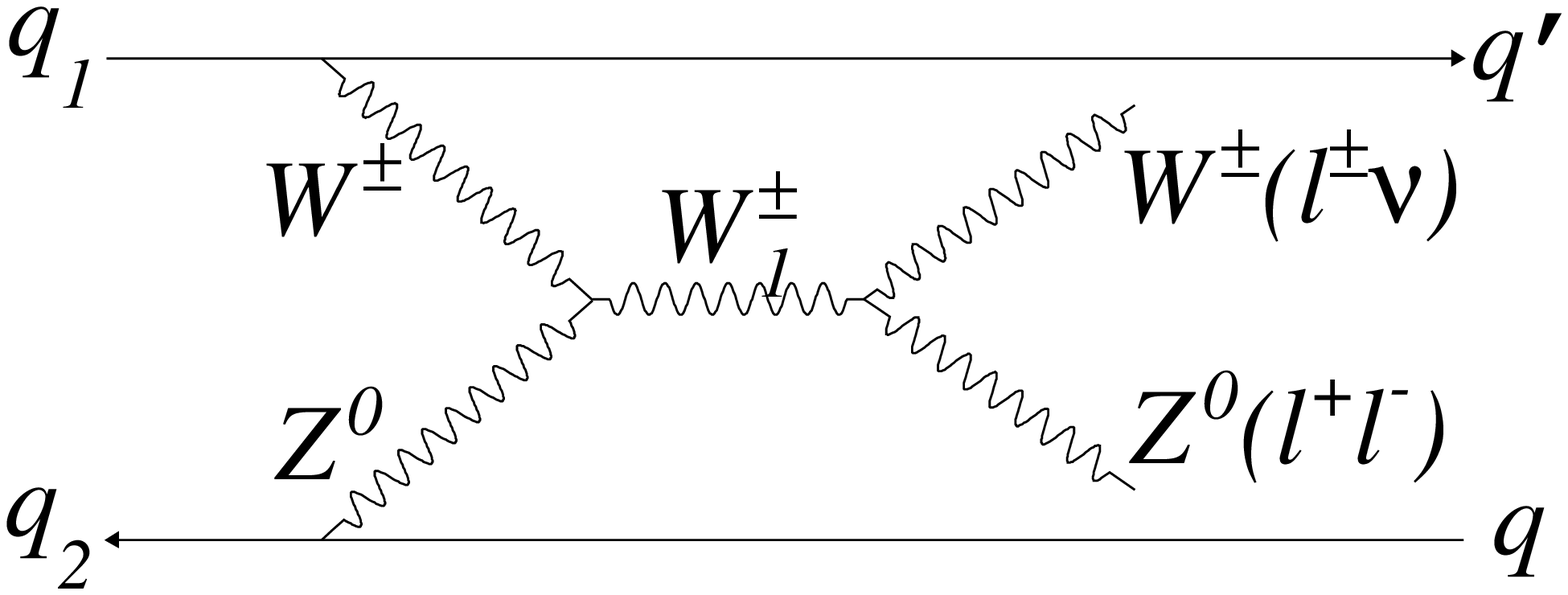}}
                    \label{fig:eps1}

Fig. 1 Feynman diagram for the process $pp\rightarrow W^{\pm}Z^0qq^\prime\rightarrow\nu3\ell qq^\prime.$
                      \end{center}
                      \end{figure}


Compared with the analysis in ref.[2], which was at the parton level, this work consider the parton showers and hadronization and studies
more backgrounds.  The hadronization and more backgrounds are
are supposed to  bring much more difficulty for data analysis and decrease the sensitivity to find the signal.
More realistic selections should be designed to suppress these backgrounds.

\section{\noindent   Signal and background production}

\subsection{\noindent Signal cross section and sample}

The generator for the signal is developed by the authors of ref.[2] based on
the matrix element calculation of the LHC process
 $pp\rightarrow {W_1}^{\pm}qq^{\prime}\rightarrow W^{\pm}Z^0qq^{\prime}
 \rightarrow \ell^{\pm}\ell^+\ell^-\nu qq^\prime$
  at a full tree level  in  the Minimal Higgsless model.
The Parton Distribution Function(PDF) CTEQ6L$^{[4]}$ is used.
   In this work the loose preselections at parton level are used to replace
    the original ones$^{[2]}$ to increase the cross section.
Quarks satisfy transverse momentum $pt_{q}>25~GeV$ and pseudorapidity $\mid \eta_q\mid <7.0.$
Leptons satisfy $pt_{\ell}>5~GeV,~\mid \eta_{\ell}\mid<2.5.$
The weighted parton level events from the generator are fed into PYTHIA for  sampling.
The initial and final state showers and
hadronization are done by PYTHIA. It should be emphasized that quarks with $\mid \eta_q\mid >4.5$
may result in some jets with $\mid \eta_j\mid <4.5$ after hadronization, but these events  were ignored
in Ref.[2]. The pseudorapidity of 4.5 is the coverage of the LHC detectors.

The cross section after the preselection, the number of events and the normalization factors
  are listed in Table. 1. The integrated luminosity is set to be 20/fb.
  The $W_1^+$ signal and $W_1^-$ signal are produced separately with 1000 events each.
  The normalization factor is $(cross~section)\times 20(fb^{-1})/events$. The uncertainties of the cross section
  are discussed in the section IV. 

For comparison, the cross section for the $W_1^{\pm}$ Drell-Yan production
 $pp\rightarrow W^{\pm}\rightarrow W_1^{\pm}Z^0\rightarrow W^{\pm}Z^0Z^0\rightarrow qq^\prime 4\ell~(\ell=e,\mu)$ is listed here, which is
$0.2585\pm0.0013\pm0.0034$, $0.1891\pm0.0008\pm0.0023$, $0.1432\pm0.0006\pm0.0029$, $0.1136\pm0.0004\pm0.0037$
and $0.0939\pm0.0003\pm0.0038$ fb for $m_{W_1^{\pm}}=550,~600,~650,~700$ and 750 GeV respectively. The first error is statistical and the second
is systematic. The integrated luminosity to find 5 $\sigma$ $W_1^{\pm}$ signal through this process is discussed in a separate paper$^{[5]}$.

 \begin{table}[htb]
\begin{center}
 Table 1{\hskip 0.4cm}Production of signal sample.
The first error is statistical and the second is systematic.
\end{center}
    \begin{center}
      \begin{tabular}{|c|ccc|ccc|} \hline
$W_1^{\pm}$ mass  & \multicolumn{3}{c|}{$W_1^+qq^\prime\rightarrow W^+Z^0qq^\prime\rightarrow \ell^+\nu \ell^+\ell^-qq^{\prime}$} &
\multicolumn{3}{c|}{$W_1^-qq^\prime\rightarrow
W^-Z^0qq^\prime\rightarrow \ell^-\nu\ell^+\ell^- qq^{\prime}$} \\
       (GeV) & cross section (fb) & events & norm &  cross section (fb) &  events & norm \\ \hline
         550 & $6.303\pm0.364\pm0.076$ & 1000 & 0.126 & $3.674\pm0.248\pm0.896$ & 1000 & 0.073 \\
         600 & $4.844\pm0.260\pm0.440$ & 1000 & 0.097 & $2.728\pm0.156\pm0.056$ & 1000 & 0.054 \\
         650 & $3.685\pm0.192\pm0.264$ & 1000 & 0.074 & $2.101\pm0.124\pm0.080$ & 1000 & 0.042 \\
         700 & $2.828\pm0.152\pm0.164$ & 1000 & 0.056 & $1.538\pm0.072\pm0.012$ & 1000 & 0.031 \\
         750 & $2.215\pm0.124\pm0.056$ & 1000 & 0.044 & $1.176\pm0.060\pm0.048$ & 1000 & 0.023 \\ \hline
      \end{tabular}
    \end{center}
  \end{table}


\subsection {\noindent Background cross sections and samples}

   There are  three leptons $(\ell=e^{\pm},\mu^{\pm})$ and two forward quarks
in the signal process. Therefore 
   the contamination to the signal process  comes from
processes including multiple leptons in the final state.
In this work, all the backgrounds but $pp\rightarrow Z^0t\overline{t}$ are generated using PYTHIA.
$pp\rightarrow Z^0t\overline{t}$ is generated using ACERMC$^{[6]}$. 
  
 In view of the final state of the signal process, the irreducible backgrounds are
$pp\rightarrow W^{\pm}Z^0\rightarrow \ell^{\pm}\nu\ell^+\ell^-$ and
  $pp\rightarrow W^{\pm}Z^0qq^\prime\rightarrow \ell^{\pm}\nu\ell^+\ell^-qq^\prime$,
hereafter $Z^0$ denotes $\gamma^*/Z^0$ for all the background channels, while 
 $Z^0$ does not include $\gamma^*$ for the signal channels though it is interfaced with PYTHIA.
The lepton components in the two backgrounds are identical to that of the signal process.
The first process is a leading order non-fusion process.
The second is a leading order fusion process, 
in which there are two forward quarks $qq^\prime$. Here fusion means
$W^{\pm}Z^0\rightarrow W^{\pm *}\rightarrow W^{\pm}Z^0$.
The cross section of the second process is about $0.06\%$ of that of the first process.
The initial and final gluon radiations will add jets to  both of them, i.e.
$pp\rightarrow W^{\pm}Z^0+ n jets$ (n=0,1,2,...; jet=q, gluon) for the non-fusion process and
$pp\rightarrow W^{\pm}Z^0qq^{\prime}+ n jets$ (n=0,...; jet=q, gluon) for the fusion process.
In Ref.[2], the non-fusion and fusion processes $pp\rightarrow W^{\pm}Z^0 2jets$ (jet=q, gluon) were discussed.

Because additional leptons  may come from
decays of quarks,
there is a  possibility that the two lepton process
 $pp\rightarrow W^+W^-\rightarrow 2\ell 2\nu$
and the corresponding fusion process
 $pp\rightarrow W^+W^-qq^\prime\rightarrow 2\ell 2\nu qq^\prime$
may construct 3 lepton states.
The second process includes $Z^0Z^0\rightarrow W^+W^-$($0.003\%$) and $W^+W^-\rightarrow W^+W^-$($99.997\%$).

If a  lepton goes beyond the detection region where  $|\eta|>2.5^{[7]}$,
the four lepton process  $pp\rightarrow Z^0Z^0\rightarrow 4\ell$
and the corresponding fusion process $pp\rightarrow Z^0Z^0qq^\prime\rightarrow 4\ell qq^\prime$
possibly become backgrounds.
If an additional lepton  comes from a quark decay,
the two  lepton process  $pp\rightarrow Z^0Z^0\rightarrow 2\ell q_i\overline{q}_i$
and the corresponding fusion process $pp\rightarrow Z^0Z^0qq^\prime\rightarrow 4\ell q_i\overline{q}_iqq^\prime$
also become backgrounds.
The second process includes 
$Z^0Z^0\rightarrow Z^0Z^0$($0.1\%$) and $W^+W^-\rightarrow Z^0Z^0$($99.9\%$). As above mentioned, in PYTHIA $Z^0$ denotes $Z^0/\gamma^*$,
the processes include $Z^0Z^0$, $Z^0\gamma^*$, $\gamma^*\gamma^*$ and their
interference. Therefore, one of the two $Z^0$ is chosen to decay into a pair of electrons or muons and another
is allowed to decay free in this work.

The multiple lepton processes that have very large cross sections are
$pp\rightarrow Z^0 b\overline{b}\rightarrow 2\ell b\overline{b}$ and
$pp\rightarrow t\overline{t}\rightarrow W^+bW^-\overline{b}\rightarrow 2\ell 2\nu b\overline{b}$
 including $q\overline{q}\rightarrow t\overline{t}$($13.9\%$) and
  $g g \rightarrow t\overline{t}$($86.1\%$).
 There are two  leptons in the final state.  
The third $\ell$ production may come from the direct or cascade decay of b quark.

The process $pp\rightarrow Z^0t\overline{t}\rightarrow 2\ell t\overline{t}$  with $t$ and $W^{\pm}$ free decay is generated
using ACERMC$^{[6]}$.  $t\overline{t}$ will produce a pair of $W^+W^-$.
 Due to the leptonic and hadronic decays of $W^{\pm}$, this process includes two leptons,
three leptons and four leptons in the final state.

The cross sections, the numbers of the events and the normalization factors for
the background samples are given in the  Table 2. It is worth
noting that all normalization factors are less than 1.
 
\begin{table}[htb]
\begin{center}
Table 2{\hskip 0.4cm}production of backgrounds.
f represents a fundamental fermion of flavour $d,~u,~s,~c,~b,~t$
\end{center}
\begin{center}
\begin{tabular}{|c|c|c|c|}\hline
Process  & cross section (fb)  & $events$ & norm. factor \\
\hline
$f\overline{f}^\prime,W^{\pm}Z^0\rightarrow W^{\pm}Z^0\rightarrow 3\ell\nu$&389.70,0.24& 6E4& 0.130\\
\hline
$f\overline{f},W^+W^-,Z^0Z^0\rightarrow W^+W^-\rightarrow2\ell2\nu$ &3295,5673,0.14&54E4& 0.332\\
\hline
$f\overline{f},W^+W^-,Z^0Z^0\rightarrow Z^0Z^0\rightarrow2\ell X$ &1676,0.54,7.4E-4&14E4& 0.240\\
\hline
$t\overline{t}\rightarrow W^+W^-b\overline{b}\rightarrow 2\ell 2\nu b\overline{b}$ &22980&2.16E6&0.213\\
\hline
$Z^0b\overline{b}\rightarrow 2\ell b\overline{b}$ &3.244E5& 2.7E7&0.240\\ 
\hline
$Z^0t\overline{t}\rightarrow 2\ell t\overline{t}$ &33.43  & 7000 &0.095\\
\hline
\end{tabular}
\end{center}
\end{table}

\section{\noindent  Event Selection}

Event selection is based on the feature of the signal process. There are three leptons with high momentum
and two hard forward quarks  at the parton level. Two of three  leptons come from
the  $Z^0$ decay and one from the $W^{\pm}$ decay. The two forward quarks are supposed to evolve into hard forward jets.
Events are accepted if
there are three leptons in the final states  and the total charge of three leptons
is equal to 1 or -1.  Each lepton satisfies $\mid \eta_\ell\mid <2.5$, $pt_{\ell}>10~GeV$.
Then the number of jets is required  to be equal to  or larger than 2. Each jet is reconstructed with cone of 
$\Delta R=\sqrt{\Delta \eta^2+\Delta \phi^2}=0.7$
and satisfies $\mid \eta_j \mid <4.5$ and $pt_{j}>25~GeV$ and $\Delta R(j,\mu)>0.3$ to suppress fake jets.

The jet reconstruction algorithm is one  provided with PYTHIA$^{[3]}$, in which
a detector grid is assumed, with the pseudorapidity range $\mid  \eta_j \mid <  \eta_{max}(4.5)$ and the
full azimuthal range each divides into 
 50 equally large bins, giving a rectangular grid; the
cell with largest $Et$ is taken as a jet initiator; a candidate jet is defined to consist of all cells
which are within $(\eta -\eta_{initiator})^2+(\phi-\phi_{initiator})^2<\Delta R^2$; the candidate jet
is accepted if the summed transverse energy $Et$ is above 7 GeV and its all cells
are removed from future consideration; the sequence is now repeated within the remaining cell of highest
$Et$ and so on.
The cut is powerful to
suppress the backgrounds $W^{\pm}Z^0\rightarrow 3\ell\nu$, $WW\rightarrow 2\ell2\nu$ and
$Z^0Z^0\rightarrow 4\ell$, because
there are no quarks  at the parton level.

\begin{center}
\epsfig{file=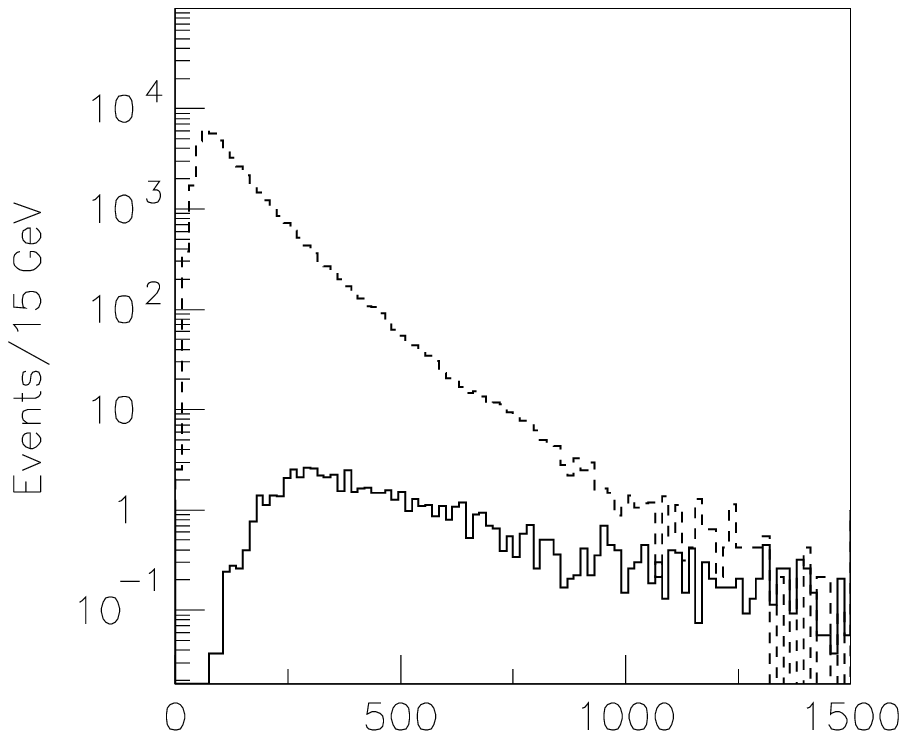 ,bbllx=25pt,bblly=418pt,%
bburx=272pt,bbury=646pt,width=8cm,height=6cm,clip=}
{\vskip -0.5cm}
{\hskip 1.4 cm} p (GeV)

{\indent}Fig. 2 The  distribution of highest momentum of the  three leptons. The solid line is for the signal.
The dashed line is for the backgrounds.
\end{center}



To reduce the backgrounds, the maximum momentum of the three leptons  is required to be larger than 150 GeV, see Fig. 2. For the signal process,
the  threshold of the momentum distribution is 150 GeV, thanks to the three leptons coming from $W_1^{\pm}$ with large mass.
Out of  the same reason, the scalar sum of pt of 3 $\ell$s and missing $Et$ is required to be larger than 700 GeV, see Fig. 3.

The reconstruction of $Z^0$ and $W^{\pm}$ using 3 $\ell$s and one missing neutrino
is a prerequisite for the signal event selection. The remaining events are required to
satisfy  $\mid m_{\mu^+\mu^-}-91.18\mid<15~GeV$ and  $W^{\pm}$ transverse mass $mt_{W^{\pm}}<200~GeV$.

 To suppress the background  $pp\rightarrow W^{\pm}Z^0$ and $Z^0b\overline{b}$ further,
the ratio of the scalar sum of jets' transverse momenta over the scalar sum of jets' momenta is larger than 0.1
and the maximum momentum of jets is larger than 200 GeV.
Finally, the transverse mass of $W^{\pm}Z^0$ denoted as $mh$ for the remaining events  is shown in Fig. 4.
Its root mean square is 226.0 GeV.
The definition of $mh$ is 
$$mh=\sqrt{Eh^2-ph^2},$$
where 
$$Eh=\sum^{3}_{i=1}E_{\ell i}+\sqrt{px_{miss}^2+py_{miss}^2},$$
$$ph=\sqrt{(\sum^{3}_{i=1}px_{\ell i}+px_{miss})^2+(\sum^{3}_{i=1}py_{\ell i}+py_{miss})^2+
(\sum^{3}_{i=1}pz_{\ell i})^2},$$
where $pi_{miss}(i=x,~y,~z)$ is minus value of sum of momenta in $i$ direction for the three leptons and the all jets.
The new definition is different from  the conventional transverse mass
$$mt=\sqrt{(\sum^{3}_{i=1}Et_{\ell i}+\sqrt{px_{miss}^2+py_{miss}^2})^2-
(\sum^{3}_{i=1}px_{\ell i}+px_{miss})^2-(\sum^{3}_{i=1}py_{\ell i}+py_{miss})^2}.$$
We call it half transverse mass temporarily, because it takes all available components
of momenta into account so that the number of components of momenta involved is larger than
that of the  transverse mass  and less than that of the mass.
For comparison, the transverse mass of $W^{\pm}Z^0$ defined as $mt$ is shown in Fig. 5.
Its root mean square is 268.5 GeV, which is worse than that of $mh$.

The number of remaining events   and  efficiency after each cut  are shown in Table 3
where $W_1^{\pm}$ mass is 700 GeV. The significance is 7.0 $\sigma$ using significance=$\sqrt{2ln(Q)},~Q=(1+N_s/N_b)^{N_{obs}}exp(-N_s)^{[8]}$,
which corresponds to 5.0 $\sigma$ for the integrated luminosity of 10.2 $fb^{-1}$.

\begin{center}
\epsfig{file=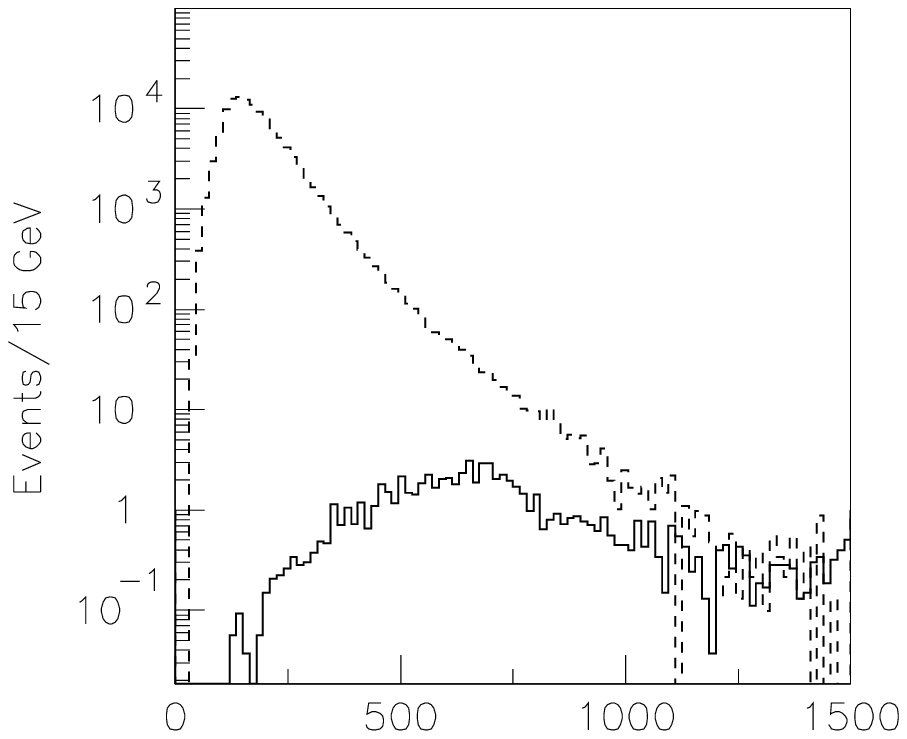 ,bbllx=25pt,bblly=418pt,%
bburx=272pt,bbury=646pt,width=8cm,height=6cm,clip=}
{\vskip -0.5cm}
{\hskip 1.4 cm}  pt(GeV)

{\indent}Fig. 3 The scalar sum of  three leptons' transverse momenta and missing pt. The solid line is for the signal.
The dashed line is for the backgrounds.
\end{center}

\begin{table}
\begin{center}
Table 3{\hskip 0.4cm} Remaining Events and Efficiency$(\%)$ for Each Cut.
The mass of $W_1^{\pm}$ is set to be 700 GeV in the table.
{\vskip 0.2cm}
\begin{tabular}{|c|c|c|c|c|c|c|c|c|}
\hline
 &        $W_1^+$& $W_1^-$& WZ& WW& ZZ& $t\overline{t}$&$Z^0b\overline{b}$ &$Z^0t\overline{t}$ \\
\hline
 $CR\times 20fb$                    & 56.6   & 30.8     & 7794.0   & 179380.0 & 33540.0 & 459600.0 & 6.48E6 &668.6  \\
                                    & 100.00$\%$ & 100.00$\%$   & 100.00$\%$   & 100.00$\%$   & 100.00$\%$  & 100.00$\%$   & 100.00$\%$ &100.00$\%$  \\
\hline
 number of                          & 54.0   & 29.5      &3260.2    & 682.3    & 1097.2    &58412.6 & 55663.2&232.0     \\
  leptons = 3                       & 95.40$\%$  & 96.00$\%$     &41.83$\%$     & 0.32$\%$     & 3.22$\%$      &12.71$\%$   & 0.86$\%$&34.70$\%$       \\
\hline
 number of                          & 46.3   & 23.7      &90.5      & 521.5    & 135.9     &25147.1 & 21114.5 &225.3    \\
 jets $\geq~2$                      & 81.80$\%$  & 77.20$\%$     &1.16$\%$      & 0.29$\%$     & 0.41$\%$      &5.47$\%$    & 0.33$\%$ 
&33.70$\%$          \\
\hline
maximum mon of                      & 45.8   & 23.4      & 50.4     & 207.6    & 67.8      &7994.7  & 5251.0 &125.3     \\
3 leps. $>150$ GeV                  & 81.0$\%$   & 76.10$\%$     & 0.65$\%$     & 0.2$\%$      & 0.20$\%$      &1.74$\%$    & 0.08$\%$ 
&18.74$\%$         \\
\hline
sum pt of 3 leps. \&                 & 24.3   & 11.1      & 2.3      & 7.3      & 1.3       &73.6    & 19.4  &7.4     \\
$pt_{missing}>700$ GeV              & 43.00$\%$  & 36.20$\%$     & 0.03$\%$    & 4.1E-3$\%$   & 3.8E-3$\%$    &0.016$\%$  & 
3.0E-4$\%$&1.10$\%$      \\
\hline
$\mid M_{\mu^+\mu^-}-M_{Z^0}\mid$   & 23.4   & 10.6      & 2.2      & 0.3      & 1.3       &7.0     & 16.1  &6.20     \\
$<15$ GeV                           & 41.30$\%$  & 34.3$\%$      &2.8E-2$\%$     & 1.9E-4$\%$   & 3.8E-3$\%$    &1.5E-3$\%$  & 
2.5E-4$\%$& 0.93$\%$      \\
\hline
 $30<M_{wt}$                         & 17.0  & 7.7     & 1.7      & 0.3      & 1.3       &2.1     & 2.4 &2.4       \\
 $<200$ GeV                         & 30.10$\%$  & 25.00$\%$     & 2.2E-2$\%$    & 1.9E-4$\%$   & 3.8E-3$\%$    &4.6E-4$\%$  & 
3.7E-5$\%$ &0.36$\%$         \\
\hline
Jets' pt/p                          & 15.2   & 7.3      &1.7       & 0        & 1.2       &2.1     & 0 &2.4       \\
$>~0.1$                             & 26.90$\%$  & 23.60$\%$     &2.2E-2$\%$    & 0$\%$        & 3.5E-3$\%$    &4.6E-4$\%$  & 
0$\%$&0.36$\%$       
\\
\hline
maximun mon of                      & 15.0   & 7.1       &1.0       & 0        & 0       &1.7     & 0  &2.0        \\
jets $>200$ GeV                     & 26.60$\%$  & 23.00$\%$& 1.3E-2$\%$&  0$\%$& 0$\%$& 3.7E-4$\%$  & 0$\%$ &0.30$\%$          \\
\hline
total                               &\multicolumn{2}{|c|}{22.1} &\multicolumn{6}{|c|}{4.7}    \\
\hline                    
\end{tabular}
{\vskip 0.1cm}
\end{center}
\end{table}

\par
{\vskip 0.2cm}

  For $m_{W_1}$ to be $550,~600,~650,~750~GeV$, the signal process can be processed with the
same criteria. The remaining events, the  significances and the integrated luminosity for $5\sigma$ signal detection are shown in Table 4.

\begin{table}[htb]
    \begin{center}
  Table 4{\hskip 0.4cm} The number of events and the significance with the integrated luminosity of $20/fb$ and the integrated luminosity for $5\sigma$ signal detection
{\vskip 0.2cm}
{\large
      \begin{tabular}{|c|cccc|cccc|c|c|} \hline
   $m_{W_1}$ &\multicolumn{4}{|c|}{number of signal}&\multicolumn{4}{|c|}{number of bg} & significance$(\sigma) with$ & lumi($fb^{-1}$) for $5\sigma$\\
       (GeV) &  &$\sigma$&misid & tot  & &$\sigma$ & misid & tot & lumi. of 20$fb^{-1}$ & signal detection \\
  \hline
   550       &  39.2 & $^{+3.9}_{-3.9}$ &$^{+0.0}_{-0.4}$ &$^{+3.9}_{-3.9}$ & 4.7 &$^{+4.7}_{-0.0}$&$^{+0.2}_{-0.0}$& $^{+4.7}_{-0.0}$  &10.9 $^{+0.98}_{-2.6}$ &4.2  $^{+3.0}_{-0.6}$          \\
 \hline
   600       &  31.4 & $^{+2.3}_{-2.3}$ &$^{+0.0}_{-0.1}$ &$^{+2.3}_{-2.3}$ & 4.7 &$^{+4.7}_{-0.0}$&$^{+0.2}_{-0.0}$& $^{+4.7}_{-0.0}$  &9.2  $^{+0.5}_{-2.1}$ &5.9  $^{+4.0}_{-0.6}$       \\
\hline
   650       &  24.6 & $^{+1.5}_{-1.5}$ &$^{+0.0}_{-0.4}$ &$^{+1.5}_{-1.6}$ & 4.7 &$^{+4.7}_{-0.0}$&$^{+0.2}_{-0.0}$& $^{+4.7}_{-0.0}$  &7.6  $^{+0.4}_{-1.8}$ &8.6  $^{+6.0}_{-0.8}$      \\
\hline
   700       &  22.1 & $^{+1.2}_{-1.2}$ &$^{+0.0}_{-0.1}$ &$^{+1.2}_{-1.2}$ & 4.7 &$^{+4.7}_{-0.0}$&$^{+0.2}_{-0.0}$& $^{+4.7}_{-0.0}$  &7.0  $^{+0.3}_{-1.6}$ &10.2  $^{+7.0}_{-0.8}$       \\
\hline
   750       &  19.7 & $^{+0.9}_{-0.9}$ &$^{+0.0}_{-0.1}$ &$^{+0.9}_{-0.9}$ & 4.7 &$^{+4.7}_{-0.0}$&$^{+0.2}_{-0.0}$& $^{+4.7}_{-0.0}$  &6.6  $^{+0.2}_{-1.5}$ &12.2 $^{+8.3}_{-0.8}$        \\
\hline

   \end{tabular}
}
    \end{center}
\end{table}

%


\section{\noindent   Systematic uncertainties}

For the analysis based on the generator level events,
the main uncertainties of $W_1^{\pm}$ discovery potential are the errors of both the signal
and background cross sections. In order to be more realistic than the generator level events,
the fake electrons and muons from jet misidentification are also estimated.

The cross section uncertainty for the signal process  includes the statistical error and the difference between
two PDFs CTEQ6L$^{[4]}$ and CTEQ6M$^{[9]}$
 listed in Table 2 as the first and second errors respectively.
The uncertainty of the background cross sections
are assumed to be $100\%$.
The uncertainties
on the  number of the signal events, the number of the background events
for luminosity of 20/fb 
are listed on Table 4.

From CMS Technical Design Report$^{[10]}$, the rates of a jet faking an electron and  
a muon due to  jet misidentification are about $6\times 10^{-4}$ and $5\time 10^{-4}$ 
respectively. To estimate the fake lepton effect on the numbers of the signal events
and the background events, the jets are misidentified by intention randomly as electrons
($50\%e^+,~50\%e^-$) and muons ($50\%\mu^+,~50\%\mu^-$)
using the rates  to re-analyze the signal process and the background processes.
The number of the signal events decreases 0.4, 0.1, 0.4, 0.1 and 0.1 for
$m_{W_1^{\pm}}=550,~600,~650,~700,~750$ GeV respectively.
The number of the background events increases 0.2 for only the process $pp\rightarrow t\overline{t}$
and keeps intact for the other processes. The uncertainities
on the numbers of the signal events and the background events are listed on Table 4. 
The uncertainties on the cross sections and the fake leptons  are added in quadrature to
the total uncertainties in Table 4.
The uncertainties on the significances for the luminosity of 20$fb^{-1}$ and the 
 luminosity to find $5\sigma$ signal
caused by the total uncertainties are also listed on Table 4.



\section{ \noindent Summary}

We have studied  discovery potential of new
 charged boson $W_1^{\pm}$ in the process
$pp\rightarrow W_1^{\pm}qq^{\prime}\rightarrow W^{\pm}Z^0qq^{\prime}\rightarrow \ell^{\pm}\nu\ell^+\ell^-qq^{\prime}(\ell=e,~\mu)$ 
based on the hadronized events with the initial and final state parton showers at the generator level. The signal generator is developed with the Minimal Higgsless
model. In addition to $pp\rightarrow W^{\pm}Z^0$, 
the five more background processes with multiple leptons are studied.
The uncertainty on the luminosity for
$5\sigma$ signal detection propagated from the uncertainties
of  cross sections is estimated. 
It is found that the major backgrounds are 
$pp\rightarrow t\overline{t}$ and $pp\rightarrow Z^0t\overline{t}$.
The integrated luminosity for 5$\sigma$
signal detection is 
between 4.2 and 12.2/fb assuming $W_1^{\pm}$ mass from  550 and 750 GeV. This integrated luminosity can be reached early in LHC.

If we only consider the backgrounds $pp\rightarrow W^{\pm}Z^0$ and $pp\rightarrow W^{\pm}Z^0qq^{\prime}$,
only 1.0 background event remains (see table 4), which means that one can find 5$\sigma$ signal of 700 GeV  using
the integrated luminosity of 5.0/fb. 
So, it can see that the other
backgrounds will bring more difficulty to
the detection of $W_1^{\pm}$.

The signal cross section decreases with the increase of $W_1$ mass.
Because we do not know how large $W_1$ mass is before analyzing data, the identical
 selection criteria is used to select the signal and background events
 for the various assumed mass of $W_1$ from 550 to 750 GeV. 
In future, when we contact with data and the LHC detector simulated events, the selection criteria can be
optimized according to practical mass of $W_1$.

This work is supported by NSFC No. 10435070, No. 10721140381, and No. 10099630, MoST No. 2007CB16101,
and CAS No. KJCX2-YW-N17 and No. 1730911111.

\begin{center}
\epsfig{file=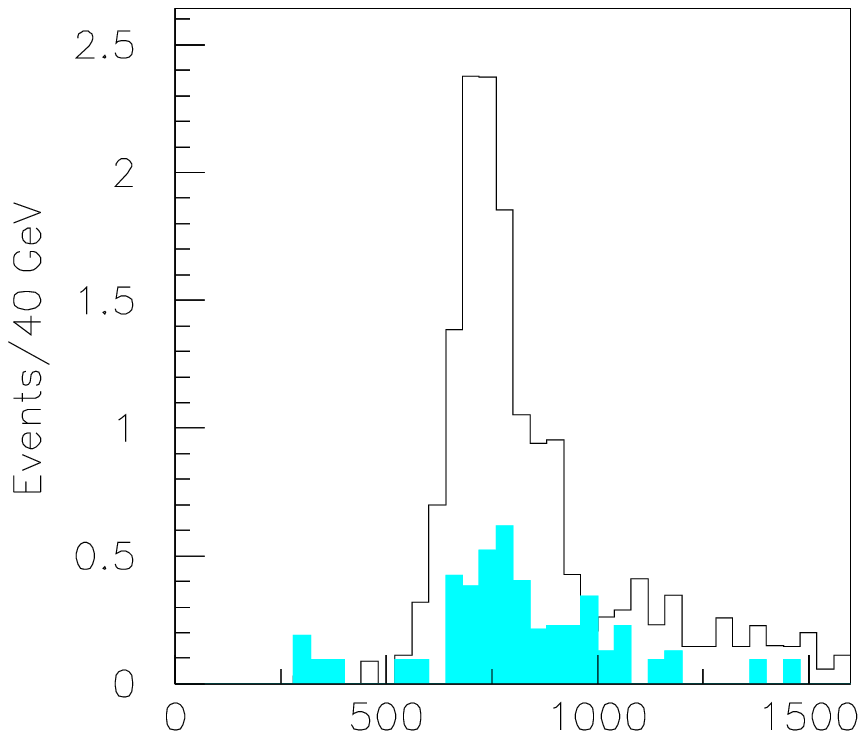 ,bbllx=25pt,bblly=418pt,%
bburx=272pt,bbury=646pt,width=8cm,height=6cm,clip=}
{\vskip -0.5cm}    
{\hskip 2cm} $mh$(GeV)

{\indent}FIG. 4 The  $W^{\pm}Z^0$ half transverse mass distribution.
The solid line is for the signal with $m_{W_1}=700~GeV$ and the integrated
luminosity of 20/fb. Its root mean square is 226.0 GeV.
The shaded area is  the backgrounds. 
\end{center}

\begin{center}
\epsfig{file=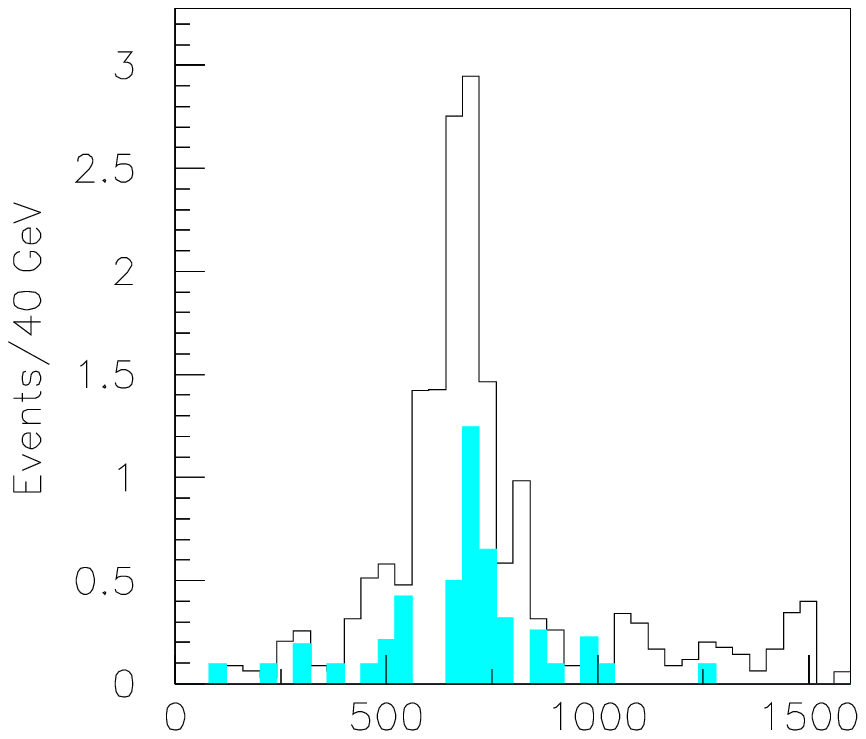 ,bbllx=25pt,bblly=418pt,%
bburx=272pt,bbury=646pt,width=8cm,height=6cm,clip=}
{\vskip -0.5cm}    
{\hskip 2cm} $mt$(GeV)

{\indent}FIG. 5 The  $W^{\pm}Z^0$  transverse mass distribution.
The solid line is for the signal with $m_{W_1}=700~GeV$ and the integrated
luminosity of 20/fb. Its root mean square is 268.5 GeV.
The shaded area is  the backgrounds. 
\end{center}

\end{document}